\documentclass[12pt,a4paper]{article}
\usepackage[utf8]{inputenc}
\usepackage{graphicx}
\usepackage{amsmath}
\usepackage{amsfonts}
\usepackage{amssymb}
\usepackage{array}
\usepackage{amscd}
\usepackage{subfig}
\usepackage{breqn}
\usepackage[margin=2cm]{geometry}
\usepackage{caption}

\title{\bf Higgs Vacuum Stability in a Mass-Dependent Renormalisation Scheme}

\vspace{5cm}

\author{\bf Alexander Spencer-Smith \\ \\
{\it ARC Centre of Excellence for Particle Physics at the Terascale, } \\
{\it School of Physics, The University of Sydney, NSW 2006 } \\
{\it e-mail: alexss@physics.usyd.edu.au}
}
\date{}

\begin{document}

\maketitle
\begin{center}

\end{center}

\begin{abstract}
\noindent

Using a physical renormalisation scheme we derive mass-dependent renormalisation group equations for the running of the Higgs quartic coupling within the Standard Model. Subsequently, we accurately take into account weak scale thresholds, resulting in a reduction of the error in the determination of the maximum $M_t$ required for absolute stability of the vacuum to 0.28 GeV. For the first time, we conclusively establish the fate of the electroweak vacuum, finding that absolute stability of the Higgs vacuum state is excluded at 99.98\% C.L. We also discuss the consequences when this new result is combined with the BICEP Collaboration's recent observation of B-mode polarisation in the cosmic microwave background, finding the Standard Model electroweak vacuum lifetime to be too short to have survived inflation. The implications for inflationary and new physics models are also discussed.
\end{abstract}

\baselineskip=16pt

\section{Introduction}

So far, all available LHC Higgs data has been found to be consistent with the predictions of the Standard Model (SM) \cite{Aad:2012tfa, Chatrchyan:2012ufa, Chatrchyan:2012jja, Chatrchyan:2013lba, Aad:2013wqa,  Aad:2013xqa}. In addition, the LHC has not detected any signals of 'beyond the Standard Model' physics \cite{Chatrchyan:2013fea, ATLAS-newphysics}, so it is reasonable to assume that the Standard Model correctly describes nature at the weak scale. With the discovery \cite{Aad:2012tfa, Chatrchyan:2012ufa} of the Higgs boson \cite{Higgs:1964ia, Higgs:1964pj, Englert:1964et, Guralnik:1964eu} with a mass of $M_h=125.9\pm 0.4$ GeV \cite{Beringer:1900zz} all the parameters of the Standard Model are now known and one can extrapolate the SM to higher energies to verify it's consistency. \\

Recently the the behaviour of the Higgs quartic coupling, $\lambda$, has been of particular interest, since it determines whether the Standard Model electroweak (Higgs) vacuum is stable, metastable or unstable \cite{Krive:1976sg, Hung:1979dn, Politzer:1978ic, Krasnikov:1978pu, Lindner:1988ww}. The previously most accurate calculations \cite{Bezrukov:2012sa, Degrassi:2012ry, Buttazzo:2013uya}  show that  the Higgs quartic coupling, $\lambda$, may become negative at energies  $\mu_I \sim 10^{11}$ GeV, if so, this means that the Higgs potential has a global minimum at large Higgs field values, $\langle h \rangle \sim 10^{17}~{\rm GeV}~ \gg v$. This minimum corresponds to the true vacuum state of the theory, with large negative energy density $\sim -10^{66}$ GeV$^{4}$. However, the previous calculations found the experimentally determined value of the Higgs mass to be $2.2$-$\sigma$ away from that required for absolute stability of the EW vacuum \cite{Buttazzo:2013uya}, which was not accurate enough to establish the fate of the vacuum conclusively. \\

In \cite{Bezrukov:2012sa, Degrassi:2012ry, Buttazzo:2013uya}, the Standard Model parameters were extrapolated to high energies by solving three-loop renormalisation group equations (RGEs) in the modified minimal subtraction scheme ($\overline{\rm MS}$), which is an example of a mass-independent renormalisation scheme in which the $\beta$-functions take an explicitly mass and renormalisation scale, $\mu$, independent form (of course, the couplings are still implicitly a function of $\mu$). The mass and scale independence of the $\overline{\text{MS}}$ RGEs makes them simpler and easier to solve than their mass-dependent counterparts. However, as a result of this mass independence, $\overline{\text{MS}}$ $\beta$-functions violate the Appelquist-Carazzone decoupling theorem \cite{Appelquist:1974tg} and one must define a sequence of effective field theories, with massive particles integrated out at their mass thresholds, and with the effective couplings of each theory suitably matched together \cite{Weinberg:1980wa}. In \cite{Buttazzo:2013uya}, the $\overline{\rm MS}$  Higgs quartic self-interaction coupling $\lambda(\mu)$, the Higgs-top Yukawa coupling $y(\mu)$ and the electroweak gauge couplings were computed by matching them with physical observables: the pole masses of the Higgs boson ($M_h$), the top quark ($M_t$), the $Z$-boson ($M_Z$), the $W$-boson ($M_W$), the $\overline{\rm MS}$ strong gauge coupling at $Z$-pole ($\alpha_3(M_Z)$), and the Fermi constant $G_F$. This matching takes place at the top quark pole-mass with 2-loop threshold corrections included. Small uncertainties arise as a result of this procedure, but since the bounds on $M_H$ and $M_t$ for absolute stability of the Higgs vacuum are so sensitive to errors in the boundary conditions it is important to reduce the uncertainties as much as possible. \\

\begin{figure}[h!]
\centering
\includegraphics[width=0.6\textwidth]{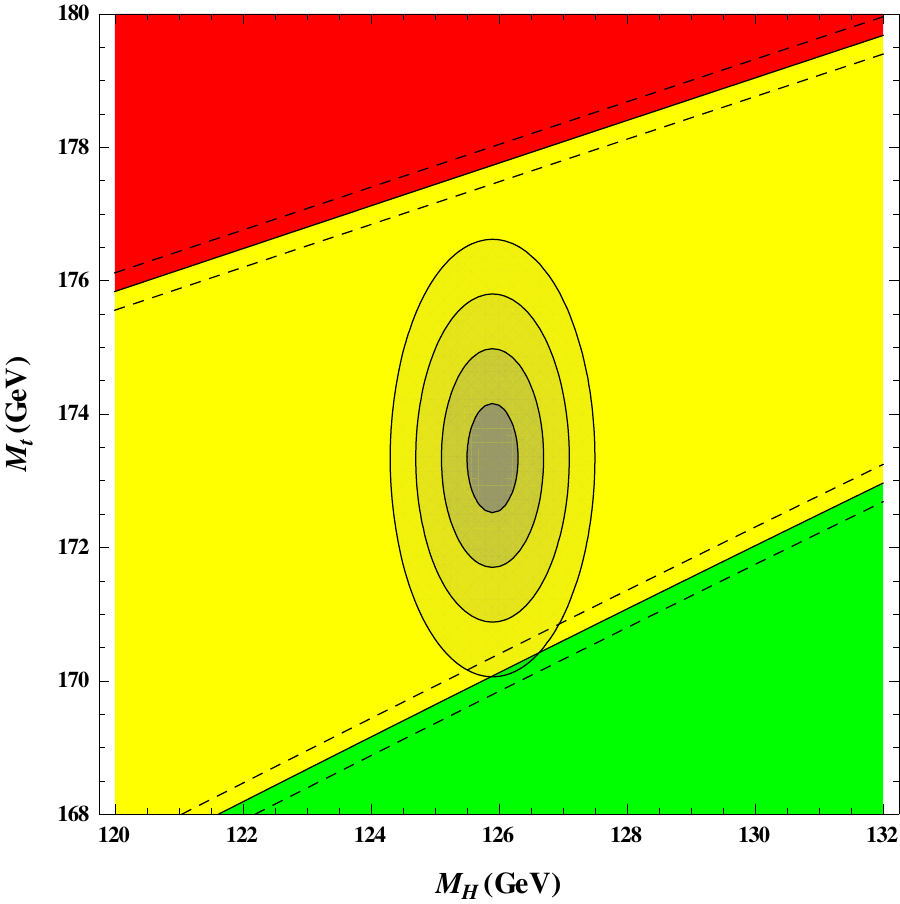}
\caption[]{Standard Model electroweak vacuum stability phase diagram in $M_t\--M_H$ plane. The colouring shows regions of EW vacuum instability (red), metastability (yellow) and stability (green). The ellipses are centred at the experimentally determined central values of $M_t$ and $M_H$ and show the $1,2,3 \ \& \ 4$-$\sigma$ error bands. The shaded regions between the dashed lines indicate the $1$-$\sigma$ bands in the determination of the maximum $M_t$ allowed for metastability (upper set of lines) and absolute stability (lower set of lines).}
\label{fig:Higgs Phase Diagram}
\end{figure}

Here we perform the evolution of all couplings for the running of $\lambda$ to one-loop in a mass-dependent scheme, which avoids the need for large matching corrections and treats threshold effects analytically. At two and three loops we use the $\overline{\text{MS}}$ RGEs from \cite{Chetyrkin:2012rz, Chetyrkin:2013wya} and second loop matching and threshold corrections from \cite{Buttazzo:2013uya}. This results in a sizeable reduction in the theoretical error associated with with the use of threshold corrections, and a reduction in the error from ignoring higher-order perturbative corrections. We now find that the largest source of uncertainty in the determination of the fate of the EW vacuum comes from the experimental error in the measurement of $M_t$. In order to ensure absolute stability of the electroweak vacuum we find the following bound on the top pole mass \cite{Kobakhidze:2014xda}
\begin{equation}
M_t < 170.16 \pm 0.28~ {\rm GeV}~.
\label{3}
\end{equation}
The upper value from the above bound, $M_t^{\rm max}=170.16 + 0.28 \ \text{GeV} = 170.44$ GeV, is $3.54$-$\sigma$ away from the central experimental value of $173.34$ GeV \cite{ATLAS:2014wva} \footnote{An alternative method for reconstructing $M_t$ using the $\overline{\text{MS}}$ mass obtained from total top pair production cross section is presented in \cite{Alekhin:2012py}. The authors of the analysis used in this work discuss uncertainties of up to $1$ GeV coming from possible differences between the measured (Monte-Carlo) and pole masses. However, even taking into account these, the total error is significantly less than the error in the analysis of \cite{Alekhin:2012py}.
We leave it to the reader to decide which is the more appropriate result.}. In other words, we find that absolute stability of the Higgs vacuum within the Standard Model is excluded at 99.98\% C.L. (one sided). The results of our improved stability analysis are shown in Figure \ref{fig:Higgs Phase Diagram}.\\

The outline for the remainder of this work is as follows: in section \ref{Schemes} we compare mass-dependent and mass-independent renormalisation schemes, then, in section \ref{Boundary Conds}, we explain the particular mass-dependent scheme we will use and the weak-scale boundary conditions for the RGEs in this scheme. Following that we perform the extrapolation of $\lambda$ and related couplings up to the Planck scale and derive bounds on the physical inputs based on the requirement of absolute stability of the electroweak vacuum in section \ref{sec:Running}. Finally, in section \ref{sec:Implications} we will discuss the physical implications of our findings, particularly those for inflationary cosmology and the consequences for new physics.

\section{Mass-dependent vs. mass-independent renormalisation schemes}
\label{Schemes}

Given an arbitrary field theory the first task at hand is to compute the Green's functions. Going beyond tree-level, one encounters divergent loop diagrams which are rendered finite through a choice of subtraction scheme.\\

In the case of an unphysical renormalisation scheme, one is purely interested in finiteness of the Greens functions, which is achieved by removing any divergent terms, and possibly some renormalisation scale, $\mu$, independent constants. Since the counterterms only contains poles and constants, the resulting $\beta$-functions take a simple mass and $\mu$ independent form, as is the case for the most commonly used mass-independent schemes, $\overline{\text{MS}}$ and $\overline{\text{DR}}$. However, we now find that massive particles do not decouple from the running of couplings. This is an artifact of the unphysical renormalisation scheme used -- the counterterms and RGEs are blind to massive particle thresholds and in general these schemes violate the Applequist-Carazzone decoupling theorem \cite{Appelquist:1974tg}. In order to satisfy the theorem and get a massive particle to decouple below it's mass threshold, $M_p$, one must set $\beta = 0$ below $M_p$ and $\beta = \beta_{\overline{\text{MS}}}$ above by hand. For more complicated theories with multiple mass thresholds one is forced to use a series of effective field theories with the massive particles integrated out at their respective threshold. The contribution of a massive particle to the RGEs is then artificially 'switched on' at threshold with a step function. Since a step function is used, this method fails to take into account the finite probability for a particle to contribute to some physical process by appearing in loop diagrams at momenta below mass threshold. \\

In practice one is able to approximately correct for these threshold effects by matching $\overline{\text{MS}}$ (or whichever unphysical scheme is being used) couplings with physical observables (e.g. pole masses). In our example we would force the low energy limit of the charge to coincide with the macroscopic electron charge. Technically this is achieved by comparing expressions for Green's functions with couplings renormalised in the on-mass-shell (OMS) scheme to those renormalised in the unphysical scheme as outlined in \cite{Weinberg:1980wa}. One is then able to invert the resultant expression and solve for the $\overline{\text{MS}}$ couplings in terms of OMS quantities and logarithms of $\mu$. Errors arise in two places -- firstly, if the Green's functions are renormalised to $n$th loop order then the matching relations will suffer errors of ($n+1$)th loop order and greater since these terms are ignored when one inverts $\overline{\text{MS}}$ couplings in terms of OMS ones. Additional errors creep in when one uses the matching formulae away from the pole masses appearing in the un-resummed logarithmic terms. Taking the Standard Model as an example and matching couplings at the top-pole we would find that reasonably large errors arise from neglecting higher order corrections in the top Yukawa, strong gauge and Higgs quartic couplings.\\

On the other hand, for a physical renormalisation scheme, such as momentum space subtraction (MOM) \cite{Georgi:1976ve} or, equivalently, an effective charge scheme, the Green's functions are directly related to a set of physical observables (e.g. pole masses and/or cross sections) at a given reference momentum through choice of the the counterterms. Then the counterterms necessarily contain finite expressions in particle masses and the renormalisation scale $\mu$, as well as divergent parts. The $\beta$-functions which follow from these will then, in general, depend explicitly on particle masses and $\mu$ (mass-dependent RGEs). Using MOM, our renormalisation conditions stipulate, for example, that the counterterms should absorb the radiative corrections to a given Green's function when the momentum $p^2 = -Q^2$, where $Q^2$ is equivalent to spacelike scale $-\mu^2$, and the resulting $\beta$-functions resulting from the renormalisation of, e.g. the two point functions, look like
\begin{equation}
\label{MOM beta}
\beta_{\rm MOM}=\beta_{\overline{\text{MS}}}\int_0^1 dz \frac{Q^2 z(1-z)}{m_p^2+Q^2 z (1-z)}.
\end{equation}
Now we find that that Appelquist-Carazzone is automatically satisfied with $\beta_{MOM} \rightarrow 0$ as $Q \rightarrow 0$ and $\beta_{\rm MOM} \rightarrow \beta_{\overline{\text{MS}}}$ as $Q \rightarrow \infty$ smoothly. By contrast with $\overline{\text{MS}}$, no matching between MOM and OMS couplings is necessary since the MOM renormalisation conditions reduce to the OMS ones when $-Q^2$ coincides with the corresponding pole mass squared. Additionally, if one uses a physical renormalisation scheme to $n$-loop order the resulting RGEs automatically encode the analytic $n$-loop threshold behaviour of the couplings in question, and also the $n$th log threshold behaviour since the logs describing the threshold behaviour are now included in the counterterms and are resummed via the renormalisation group. Alternatively, an analytic extension of the $\overline{\text{MS}}$ scheme was proposed in \cite{Brodsky:1998mf} \\

It must be stressed that, in principle, predictions for observables are renormalisation scheme independent. However, working to finite order in perturbation theory different schemes will give different predictions. In the case of QCD differences were found in the two loop running of the mass-dependent MOM $\alpha_S$ and it's $\overline{\text{MS}}$ counterpart, which could only be resolved by a rescaling of the couplings \cite{Jegerlehner:1998zg} \footnote{As discussed in \cite{Jegerlehner:1998zg} the $\overline{\text{MS}}$ coupling should be rescaled to match the MOM coupling. However, since the $\overline{\text{MS}}$ scale was already fixed for the experimental fit, the authors were forced to rescale the MOM coupling, in order to make a comparison between the two.}. This reflects the inclusion of higher order threshold effects in the RGE running of the MOM coupling as compared to the fixed order matching for the $\overline{\text{MS}}$ coupling. \\

In mass-dependent renormalisation schemes we find that the two point Green's functions  evaluate to elements of a class of log like functions \cite{Binger:2003by}, of which \eqref{MOM beta} is just one example. These arise from the appearance of Passarino-Veltman (PV) integrals \cite{Passarino:1978jh} in the evaluation of loop diagrams. The two-point functions involve the two-point PV scalar integral, $B_0(m_1,m_2, Q^2)$, which, for $-Q^2$ at a multi-particle threshold, has a pole, and develops an imaginary part for $-Q^2$ above this threshold. If we wish to use observables to determine boundary conditions on the couplings we must analytically continue our $\beta$-functions to complex $Q$ (or equivalently, real $\mu$), a prescription for the analytic continuation can be found in \cite{Binger:2003by}. To avoid the poles at multi particle thresholds one must integrate the RGEs from timelike to spacelike scales, then up the real $Q$ line far enough so that all poles are passed when one integrates back to real $\mu$. In the limit $\mu \rightarrow \infty$ all imaginary parts of the $\beta$-functions disappear, but at finite scales they contribute to the RGEs. \\

If accuracy is not the principle concern then unphysical renormalisation schemes provide the RGEs for the theory in question in their simplest mass-independent form. However, by removing some of the chief problems that plague these unphysical schemes, physical renormalisation offers a significant improvement in precision when extrapolating the couplings of a QFT to high energies. This was previously exploited to perform a high-precision analysis of gauge coupling unification in supersymmetric GUT models \cite{Binger:2003by}.  In this work we focus on a different question, namely the fate of the Standard Model electroweak vacuum, which rests upon a similarly high-precision determination of the behaviour of the Higgs quartic at high energies, under the assumption of no new physics beyond the Standard Model before the Planck scale. 

\section{Boundary conditions at the electroweak scale}
\label{Boundary Conds}

\subsection{Tree-level inputs}
\label{subsec:tree}

Tree-level inputs for our calculation are obtained from observables and are summarised in table \ref{EW inputs}.
\renewcommand{\arraystretch}{1.2}
\begin{center}
    \begin{tabular}{ | l | l | l | l | }
    \hline
    Observable & Tree-level relation to parameters & Value & Source \\ \hline
    $M_H$ & $\sqrt{2 \lambda v^2}$ & $125.9 \pm 0.4$ GeV & \cite{Beringer:1900zz} \\ \hline
    $M_t$ & $ y_t v / \sqrt{2}$ & $173.34 \pm 0.76 \pm 0.3 $ GeV & \cite{ATLAS:2014wva} \\ \hline
    $M_W$ & $g_2 v / 2$ & $80.385 \pm 0.015$ GeV & \cite{Beringer:1900zz} \\ \hline
    $M_Z$ & $v \sqrt{g_1^2 + g_2^2}/ 2$ & $91.1876 \pm 0.4$ GeV & \cite{Beringer:1900zz} \\ \hline
    $G_F$ & $(\sqrt{2}v^2)^{-1}$ & $1.1663787 \times 10^{-5} \pm 0.0000006 \ \text{GeV}^{-2}$ & \cite{Tishchenko:2012ie} \\ \hline
    $\alpha_3^{\overline{\text{MS}}}(M_Z)$ & $g_3^2 / 4 \pi$ & $0.1184 \pm 0.0007$ GeV & \cite{Bethke:2012jm} \\ \hline
    \end{tabular}
    \captionof{table}{Physical inputs at the electroweak scale with errors. We also include the tree level relation between these inputs and the couplings appearing in the Lagrangian, this also states our choice of normalisation for the couplings.}
    \label{EW inputs}
\end{center}
In table \ref{EW inputs} the first error in $M_t$ is the experimental error whilst the second is the $\mathcal{O} (\Lambda_{QCD})$ error arising from non-perturbative uncertainty in the relation between the pole and measured mass of the top. Combining these errors in quadrature gives $M_t = 173.34 \pm 0.82$ GeV.

\subsection{One-loop matching}

Keeping the discussion from the previous section in mind, we would like to renormalise all couplings relevant to the running of $\lambda$ at one-loop within a physical renormalisation scheme. MOM is the simplest choice, however, it is well known that the simplest incarnations of momentum space subtraction spoil the Slavnov-Taylor identities. To solve this problem one may use the background field method (BFM) \cite{DeWitt:1967ub,DeWitt:1967uc,DeWitt:1981,Abbott:1980hw,Abbott:1983zw}. Combining the BFM with MOM (BFMOM) restores the Slavnov-Taylor identities and reduces gauge ambiguity in the RGEs \cite{Rebhan:1985yf}. Alternatively, one can employ the Pinch technique (PT) \cite{Cornwall:1981zr} in which one works with gauge invariant effective charges. Use of PT corresponds to BFM in Feynman-'t Hooft gauge for the quantum fields \cite{Papavassiliou:1994yi}. In this work we use the one-loop mass dependent RGEs for the Standard Model gauge couplings, obtained from the PT self energies for the photon, photon-Z mixing and gluon taken from \cite{Binger:2003by}. For the running of $\lambda$ we use the PT Higgs self energy from \cite{Papavassiliou:1997pb}, obtaining the expression for the Higgs four-point counterterm using the relation given in \cite{Denner:1994xt}.  We also checked that our RGEs are gauge parameter independent using the results for the Higgs self-energy and relations between electroweak wavefunction renormalisation constants from \cite{Denner:1994xt, Denner:1991kt}. Two loop RGEs for $\alpha_S$ in BFMOM were previously derived in \cite{Jegerlehner:1998zg}. \\

Background field gauge invariance of the effective action forces the wave function renormalisation constants of the fields entering the gauge fixing to satisfy identities which differ from their usual definition in the MOM scheme, in BFMOM/PT
\begin{gather}
\delta Z_H \neq -\frac{d \Sigma^H(p^2)}{d p^2}|_{p^2=-Q^2}, \\
\delta Z_W \neq -\frac{d \Sigma^W_T(p^2)}{d p^2}|_{p^2=-Q^2}, \\
\delta Z_{ZZ} \neq -\frac{d \Sigma^{ZZ}_T(p^2)}{d p^2}|_{p^2=-Q^2},
\end{gather}
rather \cite{Denner:1994xt},
\begin{gather}
\delta Z_H = -2\delta Z_e - \frac{c_W^2}{s_W^2}\left(\frac{\delta M_W^2}{M_W^2} - \frac{\delta M_Z^2}{M_Z^2} \right)+\frac{\delta M_W^2}{M_W^2},\\
\delta Z_W  = -2\delta Z_e - \frac{c_W^2}{s_W^2}\left(\frac{\delta M_W^2}{M_W^2} - \frac{\delta M_Z^2}{M_Z^2} \right), \\
\delta Z_{ZZ}  = \delta Z_W + \left(\frac{\delta M_W^2}{M_W^2} - \frac{\delta M_Z^2}{M_Z^2} \right), \\
\delta Z_e = -\frac{1}{2}\delta Z_{AA} = \frac{1}{2}\frac{d \Sigma^{AA}_T(p^2)}{d p^2}|_{p^2=-Q^2}
\end{gather}
Since the fermionic fields do not enter the background field gauge fixing, their wavefunction renormalisation factors in BFMOM/PT are identical to those in MOM.\\

The practical implication of these differences is that we may not straightforwardly use the the OMS values $\lambda^{\rm OMS}$ and $y_t^{\rm OMS}$ as boundary conditions, since the BFMOM/PT counterterms for these couplings no longer coincide with the OMS ones when $-Q^2=M_i^2$, as one may see by comparing the SSSS-coupling in \cite{Denner:1994xt} with that in \cite{Denner:1991kt}. Instead, we must adjust couplings at one loop by taking the difference between the definitions of the wavefunction renormalisation constants, which we denote $\delta W^{(1)}(\mu)$. We find
\begin{multline}
\delta W^{(1)}(\mu)=\left(2\delta Z_H - \frac{c_W^2}{s_W^2}\left(\frac{\delta M_W^2}{M_W^2} - \frac{\delta M_Z^2}{M_Z^2} \right)+\frac{\delta M_W^2}{M_W^2}\right)^{(1)(\rm MOM)} \\ - \left(2\delta Z_H - \frac{c_W^2}{s_W^2}\left(\frac{\delta M_W^2}{M_W^2} - \frac{\delta M_Z^2}{M_Z^2} \right)+\frac{\delta M_W^2}{M_W^2}\right)^{(1)(\rm BFMOM)}
\end{multline}
In practice this correction is extremely small. Full details of the renormalisation of the electroweak, Higgs and Yukawa sectors of the Standard Model in BFMOM/PT, including the details of this calculation and of the RGEs found in the appendix, can be found in \cite{Alex}.

\subsection{Two-loop threshold corrections}

According to \cite{Weinberg:1980wa} the $n$-loop running of a coupling in $\overline{\text{MS}}$ requires matching at $n-1$-loops, so our one-loop RGEs, which automatically include one-loop threshold corrections, also encode the correct threshold behaviour for use with the two-loop $\overline{\text{MS}}$ RGEs. To use the three-loop $\overline{\text{MS}}$ RGEs we must include the two loop $\overline{\text{MS}}$ threshold correction i.e. we use only $\lambda^{(2)}(\mu)$ from the relation
\begin{equation}
\lambda^{(\overline{\text{MS}})}(\mu)=\frac{G_F}{\sqrt{2}} M_H^2 + \lambda^{(1)}(\mu)+\lambda^{(2)}(\mu)+\ldots,
\end{equation}
Similarly, we use $y_t^{(2)}(\mu)$ from the analogous expression for $y_t^{(\overline{\text{MS}})}(\mu)$, both expression are available as analytical formulae in the gauge-less limit in \cite{Degrassi:2012ry}, and as numerical expressions evaluated at $M_t$ in  \cite{Buttazzo:2013uya} . The two loop corrections to the weak gauge couplings are unknown, however, the numerical value of the one loop matching between the $\overline{\text{MS}}$ and $PT$ gauge couplings, evaluated at the Z-pole, can be extracted from table \ref{PT inputs} \cite{Binger:2003by}.
\renewcommand{\arraystretch}{1.2}
\begin{center}
    \begin{tabular}{ | l | l | l | l | }
    \hline
     & $\overline{\text{MS}}$ & $\overline{\text{PT}}_+$ & $\overline{\text{PT}}_-$ \\ \hline
    $\alpha^-1(M_Z)$ & $127.934(27)$ & $129.076(27)$ & 128.830(27) \\ \hline
    $s^2(M_Z)$ & $0.23114(20)$ & $0.23130(20)$ & $0.22973$ \\ \hline
    $\alpha_3(M_Z)$ & $0.118(4)$ & $0.130(5)$ & $0.140(5)$ \\ \hline
    \end{tabular}
    \captionof{table}{Z-pole inputs for the gauge couplings renormalised using PT. The '+' subscript indicates the coupling has been renormalised at spacelike scale, the '-' at timelike scale. Whist the numerical values are invalid as a result of new experimental results, the differences between $\overline{\text{MS}}$ and PT couplings can be used to infer the matching conditions.}
    \label{PT inputs}
\end{center}

\subsection{Boundary values at $4 M_t$}

To circumvent issues with poles at multi-particle thresholds we integrate the RGEs along a contour from renormalisation scale $\mu = M_H$ to space like scale $\mu = i M_H$ then along the complex $\mu$ line to $\mu=4 i M_t$, introducing the top quark into the two and three loop running at $M_t$, then integrating back to the timelike scale $\mu = 4 M_t$. This requires boundary values for all couplings at $\mu = M_H$. The gauge couplings were obtained from data at $M_Z$ (see table \ref{EW inputs}) then run at three loops to $M_H$. Following the above discussion $\lambda(M_H)$ is given by
\begin{equation}
\lambda(M_H) = \frac{G_F}{\sqrt{2}} M_H^2 +\delta W^{(1)}(M_H) +\lambda^{(2)}(M_H),
\end{equation}
where the first term is the tree-level expression for the quartic coupling, the second is the one-loop correction that arises from the difference between the MOM and BFMOM/PT definitions of the wavefunction renormalisation at the Higgs-pole and the third term is the two-loop $\overline{\text{MS}}$ threshold correction required for the three loop contribution to the RGEs. $y_t(M_H)$ is chosen such that, after integrating the RGEs along the contour described above we obtain the following equality at $M_t$
\begin{equation}
y_t(M_t) = \frac{\sqrt{2}M_t}{v} +\frac{1}{4}\delta W^{(1)}(M_t)+y_t^{(2)}(M_t)
\end{equation} \\

We take all couplings to be real at $M_H$ -- their imaginary parts (related to decay widths) are absorbed into the counterterms and contribute to the running of the couplings at two loops and greater (it is unclear how this was implemented in the $\overline{\text{MS}}$ analyses) \footnote{The matching conditions used in the $\overline{\text{MS}}$ scheme contain Passarino-Veltman two point integrals which develop imaginary parts at timeline scales above multi particle creation thresholds. The imaginary parts from $n$--loop matching conditions then contribute to the running of couplings at $n+2$--loop order \cite{Binger:2003by,Jegerlehner:2011mw}.}. At large scales the couplings only differ from their counterparts with the correct width by a finite imaginary part - the real parts (which are the only parts of interest here) receive the same contribution to their running as if we had taken the couplings to be complex and run with purely real RGEs.\\

The results of integrating the RGEs along the given contour are shown in table \ref{table 3}.

\begin{center}
    \begin{tabular}{ | l | l | l | l | l | l | l |}
    \hline
    Scale & $\lambda$ & $y_t$ & $g_1$ & $g_2$ & $g_3$ & $\theta_W$ \\ \hline
    $M_t$ & $0.12699 
    $& $0.98609 
    $ & $0.35695$ & $0.6532 $ & $1.25903 
    $ & 0.50019\\ \hline
    $4 M_t$ & $0.11647 
    $ & $0.93045 
    $ & $0.3595 
    $ & $0.65352 
    $ & $1.14391 
    $ I & $0.505382 
    $ \\
    \hline
    \end{tabular}
    \captionof{table}{Real parts of the boundary values for all couplings required for the running of $\lambda$, renormalised at $4 M_t$.}
    \label{table 3}
\end{center} 


\section{Running of $\lambda$ to the Planck scale}
\label{sec:Running}

Following \cite{Bezrukov:2012sa}, our condition on $M_H$ for absolute stability of the electroweak vacuum is the minimum value of $M_H$ for which 
\begin{equation}
\label{stability condition}
\lambda(\mu_0)=\beta_\lambda(\mu_0)=0
\end{equation}
for some scale $v < \mu_0 < M_{Pl}$, with all other inputs at the weak-scale set to their central values.  To obtain the condition on $M_t$ for absolute stability we use the same definition, but we now find the maximum $M_t$ that satisfies \eqref{stability condition} with $M_H$ and the other inputs fixed to their central values.
We perform the above analysis using $\lambda(\mu)$ from the RG-improved tree level potential
\begin{equation}
\label{Higgs potential}
V_{eff}(\mu,h)=\frac{\lambda(\mu)}{4}h^4.
\end{equation}
In principle a more accurate determination of these conditions can be obtained using $\lambda_{eff}(h)$ from the RG-improved two loop effective potential, however, the effective potential, renormalised in BFMOM, is unavailable. In the $\overline{\text{MS}}$ case, \cite{Buttazzo:2013uya} compared the critical value for $M_H$ obtained from the effective potential to that from the RG-improved tree-level potential and found that the difference was $0.1$ GeV. We include this in our estimate of error in the stability conditions.\\

To construct the stability phase diagram (Figure \ref{fig:Higgs Phase Diagram}) we must decide if the electroweak vacuum is stable, metastable or unstable given the inputs values for $M_H$ and $M_t$. In practice the definition \eqref{stability condition} reduces to checking if $\lambda(\mu) \ge 0$ over the range of the running, if so, the region in $M_t$--$M_H$ phase space is classified as stable. On the other hand, if we find that $\lambda(\mu) < 0$ for some $\mu$ over the running, we must then determine if this point corresponds to a metastable or unstable region in phase space. Following \cite{Isidori:2001bm} we classify a point in phase space as metastable if the current vacuum decay probability, $p < 1$. Points in phase space with $\lambda(\mu) < 0$ and $p \ge 1$ are classified as unstable. The semi-classical result for $p$ is \cite{Kobzarev:1974cp, Coleman:1977py, Callan:1977pt, Coleman:1980aw}
\begin{gather}
\label{flat decay p}
p=0.15 \frac{\mu_B^4}{H_0^4} e^{-S(\mu_B)},\\
S(\mu_B)=\frac{8\pi^2}{3|\lambda(\mu_B)|},
\end{gather}
where $H_0$ is the current Hubble rate and $S(\mu_B)$ is the action for a bounce with size $R=\mu_B^{-1}$. Since the potential in \eqref{Higgs potential} is classically scale invariant we find that bounces of any $R$ have equal action, so any $R$ will do. Radiative corrections break the scale invariance when one calculates the one-loop action for the bounce, as in \cite{Isidori:2001bm}. Formally, to use the one loop corrected $p$, one must replace $\lambda$ with the running coupling, add a correction $\Delta S$ to the action and then minimise over the scale $\mu_B$, however, approximately the same $p$ may be obtained by simply minimising $\lambda$ over the running \cite{Buttazzo:2013uya}. \\

\begin{figure}[h!]
\centering
\includegraphics[width=0.8\textwidth]{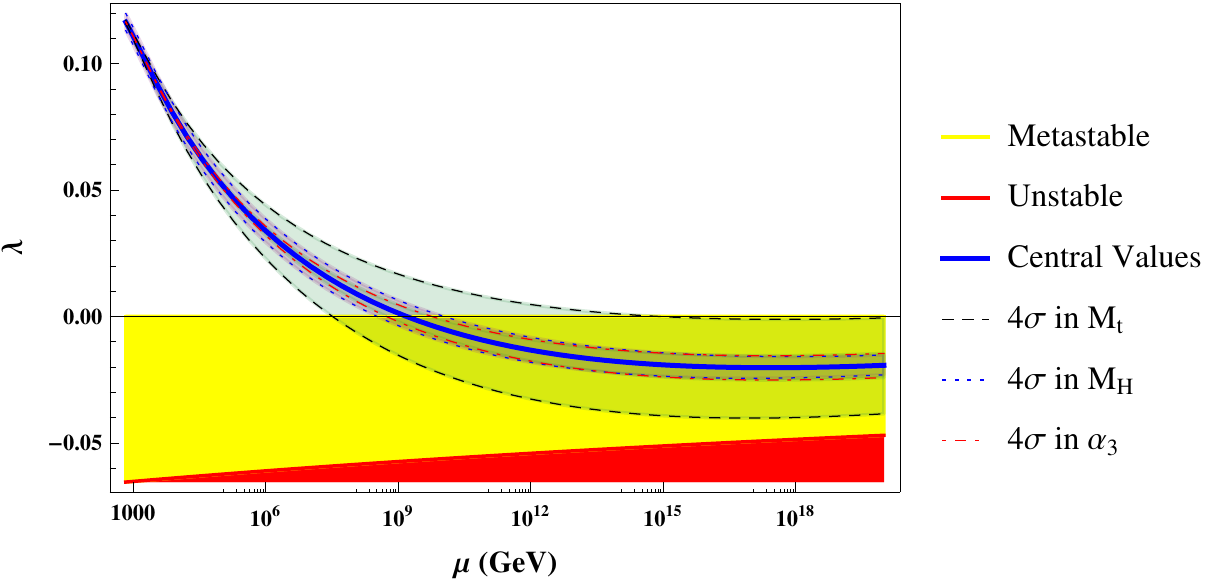}
\caption[]{Thee-loop running of the Higgs quartic coupling in the Standard Model with . The one-loop RGEs are derived in a mass-dependent scheme. The two and three loop contributions come from the $\overline{\text{MS}}$ $\beta$-functions, with effective couplings matched at two loops. The central curve shows the running for centreal values of the physical inputs, whilst the bands show a $4$-$\sigma$ variation in the inputs as labelled.}
\label{fig:Running Lambda}
\end{figure}

The three-loop running of $\lambda$ in our scheme is depicted in Figure \ref{fig:Running Lambda}. The complete set of RGEs in our renormalisation scheme may be found in appendix \ref{app:A}.  Qualitatively the running is very similar to that in the $\overline{\text{MS}}$ case - at high energies $\beta_\lambda \approx 0$ and $\lambda \approx 0$. With regard to issues concerning criticality and boundary conditions at the Planck scale, we will not add anything further than has already been discussed in \cite{Bezrukov:2012sa, Degrassi:2012ry, Buttazzo:2013uya, Holthausen:2011aa}. \\

Quantitavely we find that the instability scale $\mu_I$ (the scale at which $\lambda$ crosses zero) is lower than in the previous analyses \cite{Buttazzo:2013uya}, we now find
\begin{equation}
\log_{10}\frac{\Lambda_I}{\rm GeV}= 9.19 \pm 0.65_{M_t} \pm 0.19_{M_H} \pm 0.13_{\alpha_3}\pm 0.02_{\rm th} = 
9.19\pm 0.69,
\label{4}
\end{equation}  
where the errors are summed in quadrature in the last equality. As was the case in \cite{Buttazzo:2013uya}, we expect the scale at which the effective Higgs self-coupling (taken from the effective potential) vanishes is roughly an order of magnitude grater than that in \eqref{4}. The $\beta$-function for the running Higgs quartic coupling at the instability scale is negative and varies in the range $\beta_{\lambda}(\mu_I)\equiv \bar\beta_{\lambda}=-0.4562  \pm 0.0985$, depending on uncertainties in the input parameters. Although $\mu_I$ is a scheme \cite{Buttazzo:2013uya} and gauge \cite{DiLuzio:2014bua} dependent quantity, these ambiguities do not affect the physical implications too greatly, as we will discuss in the next section. \\

On the other hand, the value of $\lambda$ at a local minimum of the effective potential is a scheme and, in our formulation, gauge independent quantity. We find the minimum value of $\lambda$ to be lower than that obtained in the previous mass-independent analyses and we determine a new set of conditions on $M_H$ and $M_t$ for absolute stability of the electroweak vacuum. We find the condition on $M_H$ that guarantees absolute stability of the vacuum is
\begin{equation}
M_H > 132.55 \pm 1.73_{M_t} \pm 0.37_{\alpha_3} \pm 0.13_{\rm hpo} \pm
 0.1_{\rm ep}  \ \text{GeV}
\end{equation}
The experimental errors are obtained by varying the input parameters to $1$-$\sigma$ from their central values and observing the effect on the absolute stability bound. As discussed in section \ref{subsec:tree}, the experimental and non-perturbative errors in the measurement of $M_t$ are combined in quadrature. Following the prescription outlined in \cite{Degrassi:2012ry, Buttazzo:2013uya} we obtain a $0.13$ GeV uncertainty coming from higher perturbative order corrections, and $0.1$ GeV error coming from our use of the RG-improved tree-level, rather than effective, potential. In their error analysis \cite{Bezrukov:2012sa} also included an uncertainty arising from the use of the matching formulae as opposed to using the exact running. We believe that our mass-dependent scheme effectively removes this uncertainty, but since these errors were not accounted for in \cite{Degrassi:2012ry, Buttazzo:2013uya} our estimates for the uncertainties end up being roughly the same as theirs. Combining all the errors in quadrature we obtain
\begin{equation}
\label{critical MH}
M_H > 132.55 \pm 1.78 \ \text{GeV}
\end{equation}
We find a larger variation in the critical value of $M_H$ than in \cite{Buttazzo:2013uya} as a result of the increased error in the latest measurement of $M_t$ (compare  \cite{ATLAS:2014wva} with ref. [86] in \cite{Buttazzo:2013uya}). Since the error in our determination of the stability condition is greater than the error in the measurement of the Higgs mass we use the former to determine the significance of our result. The central value of the stability condition is greater than in the previous analyses, and we now find that the maximum allowed experimentally measured value for the Higgs, $M_H^{max} = 125.9+0.4 \ \text{GeV} = 126.3 \ \text{GeV}$ is $3.51$-$\sigma$ away from the value required for absolute stability, \eqref{critical MH}. \\

The experimental uncertainties in the measurement of $M_t$ provide the dominant contribution to the total error in \eqref{critical MH}, as well as being significantly larger than the error in the measurement of $M_H$. We can rephrase the condition of absolute stability as a bound on the top mass
\begin{equation}
M_t < 170.16\pm 0.22_{\alpha_3} \pm 0.13_{M_h}\pm 0.06_{\rm hpo} \pm 0.1_{\rm ep}~ {\rm GeV}~.
\label{3 at the end}
\end{equation}
Combining the errors in quadrature again, we obtain the bound \eqref{3}.
In this case, the uncertainty in the top measurement is greater than the error in the calculation of the stability condition, so we use the error in $M_t$ to work out how many $\sigma$ the maximum stability bound is from the top-pole. The upper $1$-$\sigma$ value from the above bound, $M_t^{\rm max}=170.44$ GeV, is $3.54$-$\sigma$ away from the central experimental value of $173.34$ GeV \cite{ATLAS:2014wva} \footnote{\cite{ATLAS:2014wva} discuss errors of up to $1$ GeV for total uncertainty in the relation between the MC and pole mass rather than the $0.3$ GeV error taken from \cite{Degrassi:2012ry} and used here. Accounting for this larger error we find our stability condition to be $2.3$-$\sigma$ from $M_t$. Note that if the same error were applied to the previous analyses then the statistical significance of those stability conditions would shift by a corresponding factor.}. Thus, we arrive at the most important new result of our work -- absolute stability of the Higgs vacuum within the Standard Model is excluded at 99.98\% C.L. (one sided). \\

\section{Physics implications}
\label{sec:Implications}

\subsection{Inflation}

Very recently there has been a renewal of interest in the cosmological aspects of the vacuum stability problem, with the BICEP Collaboration's announcement of their discovery of B-mode polarisation in the Cosmic Microwave Background (CMB) \cite{Ade:2014xna}. These polarisation modes are widely attributed to primordial gravitational waves. BICEP measures a tensor-to-scalar amplitude ratio of \cite{Ade:2014xna}
\begin{equation}
r=0.2^{+0.07}_{-0.05}~,
\label{5}
\end{equation}
with which one can infer the rate of inflation: 
\begin{equation}
H_{\rm inf}\approx  10^{14}~{\rm GeV}~.
\label{6}
\end{equation}
Given such a large rate of inflation, and assuming a simple model in which the inflaton does not couple to the Higgs (the case of Higgs-inflaton interations will be discussed later), the transition from a metastable electroweak Higgs vacuum to the true vacuum (with large negative energy density) is dominated by the Hawking-Moss (HM) instanton \cite{Hawking:1981fz}. An analytical expression for the decay rate of the electroweak vacuum via the HM instanton was calculated and implications for inflationary models discussed in \cite{Kobakhidze:2013tn}. The probability that the electroweak vacuum, $h=v$, decays during `visible' inflation,  $N_e=\tau_{\rm inf}H_{\rm inf}\approx 60$, via the HM instanton is $\left(1-e^{-p}\right)$, where
\begin{equation}
p=N_e^4\exp\left\lbrace \frac{\pi^2\bar\beta_{\lambda}}{2e}\frac{\mu_I^4}{ H^4_{\rm inf}} \right\rbrace~,
\label{7}
\end{equation}
and $\bar\beta_{\lambda}$ is $\beta_\lambda$ evaluated at the instability scale $\mu_I$ (the expression \eqref{7} was also derived using stochastic arguments in \cite{Enqvist:2014bua}). Given the large inflationary scale \eqref{6} and the SM instability scale \eqref{4}, the numerical value of the decay probability via the HM instanton was calculated in \cite{Kobakhidze:2014xda} and found to be $p_{\rm SM} \approx N_e^4$. Even though there is up to a two order-of-magnitude gauge dependence in the instability scale used to calculate $p$ \cite{DiLuzio:2014bua}, since the rate of inflation \eqref{6} is so large it turns out that the result for $p_{\rm SM}$ is robust. Consequently, the measured value of $M_t$ must be very close to the stability bound \eqref{3}, otherwise vacuum decay prevents simple inflation from proceeding. In fact, repeating our analysis from section \ref{sec:Running}, with the flat spacetime decay probability \eqref{flat decay p} replaced with the decay probability during inflation \eqref{7}, we find that, in order to ensure inflation proceeds, the top quark mass must satisfy the condition
\begin{equation}
M_t < 170.54 \ \text{GeV}
\end{equation}
which is $3.4$-$\sigma$ away from the experimental value (table \ref{EW inputs}). Given that the required value of the top-pole is so far from the experimentally measured value, we can conclude that the electroweak vacuum could not have survived simple inflation if one assumes no new physics beyond the SM up to the Planck scale.  \\

It has been argued that, within the multiverse picture of eternal inflation \cite{Guth:2007ng}, only patches of universe with electroweak vacuum initial conditions survive due to the collapse of AdS bubbles formed if vacuum decay occurs \cite{Espinosa:2007qp}, which amounts to a process of 'cosmological selection'. However, as was found in \cite{Kobakhidze:2013tn}, Higgs vacuum decay via the HM instanton occurs quickly enough for inflation to cease globally and not only within one Hubble volume, and, again, we arrive at the conclusion from our previous paragraph.\\

Now we discuss the implications of our new results for non-minimal models of inflation. It has been suggested that non-minimal inflationary models such as those with inflaton-Higgs interactions, or non-minimal Higgs-gravity interactions (that induce a large effective mass for the Higgs boson during inflation), may suppress Higgs vacuum decay, even with a large inflationary rate \cite{Lebedev:2012sy, Busoni:2014sya, Fairbairn:2014zia, Kehagias:2014wza}. However, these interactions only suppress the mechanism whereby the Higgs vacuum decays due to superhorizon quantum fluctuations during inflation \cite{Espinosa:2007qp}. In fact, in inflationary models with inflaton-Higgs or inflaton-gravity coupling, the dominant mechanism of Higgs vacuum decay is the Coleman-de Lucia (CdL) instanton \cite{Kobakhidze:2013tn}. Using our new results \eqref{4} and \eqref{3}, one finds that, again, the decay probability within these models is large enough to prevent inflation from proceeding \cite{Kobakhidze:2014xda}. In addition, recent work suggests that the survival probability of patches which decay via super horizon fluctuations may be lower than previously thought \cite{Hook:2014uia}.  \\

All the simple models of Higgs-inflation \cite{Bezrukov:2007ep} are now excluded as well, since these models require absolute stability of the electroweak vacuum, which we have found to be excluded at $3.5$-$\sigma$. Uncertainties in the previously most accurate NNLO $\overline{\text{MS}}$ analysis of Higgs inflation within the SM were slightly too large to rule the model out in it's simplest incarnation \cite{Salvio:2013rja}. An alternative argument which also rules out simple Higgs-inflation was presented in \cite{Fairbairn:2014nxa}. \\

If we make the assumption that the BICEP measurements are indeed produced by primordial gravitational waves then, no matter the inflationary model, it is very unlikely the Standard Model Higgs vacuum could have survived a period of inflation with such a high rate.
\subsection{New physics}

Of course, as was the case for the non-minimal inflationary models discussed above, one may introduce new physics which alters the shape of the Higgs potential at large field values, as long as this new physics comes into play below the instability scale,  $M_{\rm np} \lesssim \mu_I\sim 10^{9}$ GeV. If we are to take the previous discussion seriously then any new physics model must ensure absolute stability of the Higgs vacuum. \\

Generically, new physics will affect the Higgs potential through the running of $\lambda$. As in the standard model, the couplings between new physics and the Higgs will appear in the $\beta$-function for lambda. The sign with which these couplings appear, as well as any threshold effects, will determine the effect of the new physics upon the Higgs potential. Then one may turn the vacuum stability problem on it's head and use the requirement of Higgs vacuum stability to rule out simple extensions of the Standard Model. \\

Simple situations in which the introduction of a scalar singlet stabilises the Higgs potential have been widely discussed \cite{Lebedev:2012zw, EliasMiro:2012ay}. However, since we already require physics beyond the Standard Model to explain neutrino masses and dark matter, it is reasonable to expect that this new physics may also be responsible for the stabilisation of the Higgs vacuum. The question of vacuum stability within models of dark matter and GUT models has been discussed in \cite{Gonderinger:2009jp, Kadastik:2011aa, Gonderinger:2012rd, Chen:2012faa, Kannike:2012pe, Anchordoqui:2012fq, Allison:2012qn, Khan:2012zw, Chao:2012xt, Goudelis:2013uca, He:2013tla}. Vacuum stability in various models of neutrino mass generation were performed in \cite{Casas:1999cd, Rodejohann:2012px, Chakrabortty:2012np, Gogoladze:2008ak, Chen:2012faa, He:2012ub, Gogoladze:2008gf, Arina:2012fb, Chun:2012jw, Chao:2012mx, Dev:2013ff} and a general analysis of the effect of new physics upon Higgs vacuum stability, using neutrino masses as a case study, was performed in \cite{Kobakhidze:2013pya}.

\section{Conclusion}

We have improved upon the the mass-independent  $\overline{\text{MS}}$ analyses of Higgs vacuum stability by recalculating the running of the Higgs quartic coupling, $\lambda$, in the mass-dependent BFMOM/PT renormalisation scheme. We found that the critical value of $M_t$ required for absolute stability of the electroweak vacuum is lower than previously thought, and the critical value of $M_h$ is greater than previously thought. Taking the most significant of these two conditions, we now find that absolute stability of the Higgs vacuum to be excluded at 99.98\% C.L. -- we may now say conclusively that the Standard Model Higgs vacuum lives in a region of metastability. We combined these new results with the recent BICEP measurements of B-mode polarisation in the Cosmic Microwave Background and, under the assumption that this polarisation is due to primordial gravitational waves, were able to rule out a wide range of currently popular inflationary models. Finally, we discussed some new physics scenarios, motivated by the requirement of absolute stability of the electroweak vacuum.

\section{Acknowledgements}

I would like to thank Archil Kobakhidze for continuing help and advice on many aspects of this work. Thanks also to Stanley Brodsky for helpful correspondence, particularly regarding issues surrounding the PT couplings at the weak scale. This work was partially supported by the ARC.

\appendix


\section{Passarino-Veltman two-point functions}
\label{app:B}

Following \cite{Denner:1991kt}, The PV two-point function evaluates to
\begin{dmath}
B_{0}(m_{0},m_{1},p_{10}^2) = \Delta -
\int^{1}_{0} dx \log\frac{[p^{2}_{10}x^{2}-x(p^{2}_{10}-
m^{2}_{0}+m^{2}_{1})+m^{2}_{1}-i\varepsilon ]}{\mu^{2}}+O(\epsilon) \\
=\Delta + 2 - \log\frac{m_{0}m_{1}}{\mu^{2}}
+\frac{m^{2}_{0}-m^{2}_{1}}{p^{2}_{10}}\log\frac{m_{1}}{m_{0}}
-\frac{m_{0}m_{1}}{p^{2}_{10}}\left(\frac{1}{r}-r\right)\log r
+O(\epsilon),
\label{B0}
\end{dmath}
where
\begin{equation}
\Delta = \frac{2}{4-d}-\gamma_E+\log{4\pi}=\frac{2}{\epsilon}-\gamma_E+\log{4\pi},
\end{equation}
and $r$ and $\frac{1}{r}$ are determined from
\begin{equation}
x^{2}+\frac{m^{2}_{0}+m^{2}_{1}-p^{2}_{10}- i \varepsilon
}{m_{0}m_{1}}\,x+1=\left(x+r\right)\left(x+\frac{1}{r}\right) .
\end{equation}
We define the 'reduced' PV two-point function as
\begin{dmath}
\overline{B_{0}}(m_{0},m_{1},p_{10}^2) =  -
\int^{1}_{0} dx \log[p^{2}_{10}x^{2}-x(p^{2}_{10}-
m^{2}_{0}+m^{2}_{1})+m^{2}_{1}-i\varepsilon ]+O(\epsilon) \\
=2 - \log{m_{0}m_{1}}
+\frac{m^{2}_{0}-m^{2}_{1}}{p^{2}_{10}}\log\frac{m_{1}}{m_{0}}
-\frac{m_{0}m_{1}}{p^{2}_{10}}\left(\frac{1}{r}-r\right)\log r
+O(\epsilon).
\label{B0}
\end{dmath}
In our $\beta$-functions we use the quantities
\begin{equation}
\overline{\text{DB}_0}(m_0, m_1, \mu^2) = \mu \frac{\partial \overline{\text{B}_0}(m_0, m_1, \mu^2)}{\partial \mu} .
\end{equation}

\section{One-loop mass-dependent RGEs}
\label{app:A}

To ensure gauge invariance of the counterterms we include tadpole diagrams in the renormalisation of the one and two point functions \cite{Fleischer:1980ub}. As a result, some counterterms contain the one point Passarino-Veltman functions, $A_0(m)$. However, since the tadpole diagrams are independent of external momentum, they do not contribute to our $\beta$-functions (neither do seagull diagrams for the same reason). The one-loop mass-dependent strong and electroweak gauge coupling RGEs, together with the relevant self-energies, are rather lengthy but complete expressions may be found in \cite{Binger:2003by}. All of the following $\beta$-functions approach their $\overline{\text{MS}}$ counterparts as $\mu \rightarrow \infty$.

\subsection{Higgs quartic coupling}

\begin{dmath}
-32\pi^2 \mu \frac{d \lambda}{d\mu}= \lambda \left(12 y_t^2\overline{\text{DB}_0}(M_t, M_t, \mu^2) - 3 (g_1^2+g_2^2)\overline{\text{DB}_0}(M_Z, M_Z, \mu^2) -6 g_2^2\overline{\text{DB}_0}(M_W, M_W, \mu^2)  \right) + 24\lambda^2 \left(\frac{1}{6}\overline{\text{DB}_0}(M_W, M_W, \mu^2) + \frac{1}{12}\overline{\text{DB}_0}(M_Z, M_Z, \mu^2) + \frac{3}{4}\overline{\text{DB}_0}(M_H, M_H, \mu^2)  \right) + \frac{3}{8}(g_1^2+g_2^2)^2 \overline{\text{DB}_0}(M_Z, M_Z, \mu^2) + \frac{3}{4}g_2^4 \overline{\text{DB}_0}(M_W, M_W, \mu^2) -6y_t^4\overline{\text{DB}_0}(M_t, M_t, \mu^2)
\end{dmath}

\subsection{Top Yukawa coupling}

\begin{dmath}
-32\pi^2 \mu \frac{d y_t}{d\mu}= y_t \left( y_t^2 \left(3 \overline{\text{DB}_0}(M_t, M_t, \mu^2) + \frac{1}{2}\left(1-\frac{M_t^2-M_H^2}{\mu^2}\right)(\overline{\text{DB}_0}(M_t, M_Z, \mu^2) +\overline{\text{DB}_0}(M_t, M_H, \mu^2)) + \frac{1}{2}\left(1+\frac{M_W^2}{\mu^2}\right)\overline{\text{DB}_0}(M_b, M_W, \mu^2) \right) 
- g_2^2 s_W^2 \left( \frac{s_W^2}{c_W^2}\frac{4}{9}\overline{\text{DB}_0}(M_t, M_Z, \mu^2) +\frac{4}{9}\overline{\text{DB}_0}(M_t, 0, \mu^2) 
- \frac{1}{2}\left( \overline{\text{DB}_0}(M_t, M_Z, \mu^2) -\overline{\text{DB}_0}(M_t, M_H, \mu^2) \right)   \right)
- g_2^2 s_W^2 \left( \frac{\left(\frac{1}{2}-\frac{2}{3}s_W^2\right)^2}{s_W^2 c_W^2}\overline{\text{DB}_0}(M_t, M_Z, \mu^2) +\frac{4}{9}\overline{\text{DB}_0}(M_t, 0, \mu^2)  +\frac{1}{2s_W^2} \overline{\text{DB}_0}(M_t, M_W, \mu^2) \right)
+ 2g_2^2 s_W^2 \left( \frac{8}{3}\frac{\left(\frac{2}{3}s^2_W-\frac{1}{2}\right)}{c_W^2}\overline{\text{DB}_0}(M_t, M_Z, \mu^2) +\frac{16}{9}\overline{\text{DB}_0}(M_t, 0, \mu^2)  \right)
-2g_2^2 \overline{\text{DB}_0}(M_W, M_W, \mu^2)
-(g_1^2+g_2^2)\overline{\text{DB}_0}(M_Z, M_Z, \mu^2)
-8g_3^2\overline{\text{DB}_0}(M_t, 0, \mu^2) 
-\frac{y_t v}{2\sqrt{2}}\frac{\partial}{\partial \mu}\left( + \frac{y_t^2}{2}\left(1-\frac{M_t^2-M_H^2}{\mu^2}\right)(\overline{\text{DB}_0}(M_t, M_Z, \mu^2) +\overline{\text{DB}_0}(M_t, M_H, \mu^2)) + \frac{y_t^2}{2}\left(1+\frac{M_W^2}{\mu^2}\right)\overline{\text{DB}_0}(M_b, M_W, \mu^2) - g_2^2 s_W^2 \left( \frac{s_W^2}{c_W^2}\frac{4}{9}\overline{\text{DB}_0}(M_t, M_Z, \mu^2) +\frac{4}{9}\overline{\text{DB}_0}(M_t, 0, \mu^2)   \right)
- g_2^2 s_W^2 \left( \frac{\left(\frac{1}{2}-\frac{2}{3}s_W^2\right)^2}{s_W^2 c_W^2}\overline{\text{DB}_0}(M_t, M_Z, \mu^2) +\frac{4}{9}\overline{\text{DB}_0}(M_t, 0, \mu^2)  +\frac{1}{2s_W^2} \overline{\text{DB}_0}(M_t, M_W, \mu^2) \right)
+ 2g_2^2 s_W^2 \left( \frac{8}{3}\frac{\left(\frac{2}{3}s^2_W-\frac{1}{2}\right)}{c_W^2}\overline{\text{DB}_0}(M_t, M_Z, \mu^2) +\frac{16}{9}\overline{\text{DB}_0}(M_t, 0, \mu^2)  \right) 
- \frac{y_t^2}{2}\left( \overline{\text{DB}_0}(M_t, M_Z, \mu^2) -\overline{\text{DB}_0}(M_t, M_H, \mu^2) \right)  \right)
\right)
\end{dmath}

\end{document}